\documentclass[conference]{IEEEtran}
\IEEEoverridecommandlockouts
\usepackage{cite}
\usepackage{amsmath,amssymb,amsfonts}
\usepackage{algorithmic}
\usepackage{graphicx}
\usepackage{textcomp}
\usepackage{xcolor}
\usepackage{amsthm}
\usepackage{comment}
\usepackage{graphicx}
\usepackage[caption=false,font=footnotesize]{subfig}
\def\BibTeX{{\rm B\kern-.05em{\sc i\kern-.025em b}\kern-.08em
    T\kern-.1667em\lower.7ex\hbox{E}\kern-.125emX}}
\begin{document}

\newtheorem{theorem}{Theorem}
\newtheorem{proposition}{Proposition}
\newtheorem{lemma}{Lemma}
\newtheorem{corollary}{Corollary}
\newtheorem{definition}{Definition}
\newtheorem{remark}{Remark}

\def\re{\color{red}}
\def\jerry{\color{violet}}

\title{On the Communication-Sensing Tradeoff in DMT-Optimal Space-time-Coded ISAC Systems\\
}

\author{
\IEEEauthorblockN{
Wei-Chen Chen,
Yu-Chih Huang, 
and Ming-Chun Lee
}

\IEEEauthorblockA{
\textit{Institute of Communications Engineering, National Yang Ming Chiao Tung University, Hsinchu, Taiwan} \\
Email: wayne1009.ee13@nycu.edu.tw, jerryhuang@nycu.edu.tw, mingchunlee@nycu.edu.tw
}

}
\maketitle
\begin{abstract}
Integrated sensing and communications (ISAC) has been extensively studied from the perspective of communication--sensing performance tradeoffs. In MIMO communication systems, the diversity--multiplexing tradeoff (DMT) fundamentally characterizes the tradeoff between communication reliability and multiplexing gain, and DMT-optimal space-time codes achieve this optimal tradeoff without instantaneous channel state information at the transmitter (CSIT), making them particularly attractive for ISAC scenarios where accurate CSIT is difficult to acquire or maintain. 
In this work, we propose an analytical framework for characterizing the sensing behavior and communication--sensing gain tradeoff of DMT-optimal space-time codes in ISAC systems. Under a single-target detection model, we propose the sensing diversity order as the high-SNR sensing reliability metric. We further show that a class of DMT-optimal space-time codes, exemplified by Perfect codes, eliminates sensing blind spots for all nonzero codewords, thereby achieving joint DMT–sensing-diversity (DMTS) optimality.
Furthermore, using the Perfect code as a representative DMT-optimal construction, we show that its isotropic average transmit covariance limits the achievable sensing gain when the target angular region is narrow. We therefore develop a full-rank spatial precoding framework and a low-complexity surrogate precoder that preserve joint DMTS optimality while shaping the transmit beampattern. Simulation results show that, for a target angular region of $\theta\in[-\pi/10,\pi/10]$, the proposed design achieves approximately $3$ dB sensing-gain improvement over the unprecoded Perfect code.
\end{abstract}

\begin{IEEEkeywords}
ISAC, space-time coding, diversity--multiplexing tradeoff, sensing diversity,
full-rank precoding, communication--sensing tradeoff.
\end{IEEEkeywords}

\section{Introduction}

Integrated sensing and communications (ISAC) has emerged as a key enabling technology for future sixth-generation (6G) wireless networks \cite{liu2022integrated,gonzalez2024integrated}. ISAC enables the efficient reuse of spectrum, hardware, and signal-processing resources while supporting both communications and sensing. Existing studies have characterized fundamental communication--sensing tradeoffs, including the CRB--rate region \cite{xiong2023fundamental}, the Pareto boundary for MIMO ISAC systems \cite{hua2023mimo}, and the informative subspace structure of Pareto-optimal signaling \cite{taricco2026subspace}. To balance communication and sensing performance, transmit covariance optimization and joint beamforming have been extensively studied \cite{liu2020joint,hua2023mimo}, including WMMSE- and SCA-based designs \cite{liao2025design} and deep unfolding approaches \cite{chiu2025deep}. Related studies have also considered imperfect or statistical CSIT through outage-based beamforming \cite{bazzi2023outage} and MSE-based optimization \cite{he2024mse}. Despite these advances, the implications of such designs for the DMT and sensing reliability of space-time coded transmissions remain unclear.

Space-time coding exploits the spatial and temporal degrees of freedom of multiple-input multiple-output (MIMO) channels by jointly encoding information across transmit antennas and channel uses. The resulting diversity and multiplexing gains are fundamentally characterized by the diversity--multiplexing tradeoff (DMT) \cite{zheng2003diversity}. In particular, properly constructed full-rate space-time codes satisfying the non-vanishing-determinant (NVD) property can achieve the optimal DMT \cite{elia2006explicit,oggier2007cyclic}, with the Perfect code being a representative construction \cite{oggier2006perfect}. Importantly, such DMT optimality can be achieved without instantaneous channel state information at the transmitter (CSIT), making space-time coding particularly attractive for ISAC systems where accurate CSIT is difficult to acquire or maintain.

More recently, the communication DMT of MIMO ISAC systems under sensing constraints was investigated in \cite{du2026sensing}. Nevertheless, the fundamental sensing behavior of conventional DMT-optimal space-time codes remains largely unexplored. Furthermore, incorporating sensing-oriented spatial precoding introduces an additional challenge, as rank-deficient precoding generally cannot preserve the underlying optimal DMT. These considerations motivate the investigation of inherent sensing guarantees of DMT-optimal space-time codes and full-rank precoding strategies that preserve communication reliability while improving sensing performance.

In this work, we establish an analytical framework for characterizing the sensing behavior of DMT-optimal space-time codes and investigate the resulting communication--sensing gain tradeoff. The main contributions are summarized as follows:

\noindent {\bf (1)} We establish a unified analytical framework for space-time-coded ISAC systems and derive the miss-detection probability for arbitrary space-time codewords under a single-target detection model. We further characterize the high-SNR sensing behavior, showing that the maximum sensing diversity order of one is achieved in the absence of sensing blind spots.

\noindent {\bf (2)} We establish a connection between the algebraic properties of space-time codes and their sensing diversity. In particular, we show that the NVD property of the considered space-time code constructions eliminates sensing blind spots, enabling joint DMT--sensing-diversity (DMTS) optimality. Using the Perfect code as a representative example, we further demonstrate that its isotropic average transmit covariance can lead to sensing-gain loss over narrow target angular regions.

\noindent {\bf (3)} We develop a full-rank spatial precoding framework that preserves joint DMTS optimality while enabling spatial sensing-gain shaping. We characterize the tradeoff between communication coding gain and sensing gain and propose a low-complexity surrogate precoder for narrow sensing regions.

\noindent {\bf (4)} Numerical simulations validate the theoretical sensing-diversity analysis and illustrate the communication--sensing gain tradeoff of the proposed precoding framework. In particular, for a target angular region of $\theta\in[-\pi/10,\pi/10]$, the proposed design achieves approximately $3$ dB of sensing-gain improvement over the unprecoded Perfect code.



\section{System Model and Performance Metrics}
This section introduces the space-time-coded MIMO ISAC model and the proposed performance metrics.

\subsection{ISAC System Model}

We consider a narrowband monostatic MIMO ISAC system, where an
$N_t$-antenna transmitter simultaneously serves a communication
receiver and senses a target using the same space-time waveform.
The transmitter and sensing receiver are assumed to be colocated,
whereas the communication receiver may be located separately. This
model represents, for example, an ISAC base station that simultaneously
provides wireless connectivity and monitors targets within its
surrounding environment.

During each transmission block of $T$ channel uses, the transmitter
sends a space-time codeword $\mathbf{X}\in\mathbb{C}^{N_t\times T}$
selected from a codebook $\mathcal{C}$, with
$\mathbb{E}[\|\mathbf{X}\|_F^2]=T$, where the expectation is taken
over the transmitted codewords.

The communication channel follows a quasi-static block-fading model, where $\mathbf{H}_c\in\mathbb{C}^{N_c\times N_t}$ remains constant within each space-time coding block and varies independently across blocks. The entries of $\mathbf{H}_c$ are assumed to be independent and identically distributed as $\mathcal{CN}(0,1)$, corresponding to Rayleigh fading. No CSIT is assumed.
The communication received signal is
\begin{equation}
    \mathbf{Y}_c
    =
    \sqrt{\rho_c}\mathbf{H}_c\mathbf{X}
    +\mathbf{N}_c,
    \label{eq:communication_model}
\end{equation}
where $\rho_c$ denotes the communication SNR and the entries of
$\mathbf{N}_c$ follow $\mathcal{CN}(0,1)$.

For sensing, we consider a single far-field target located at angle
$\theta$. The target is assumed to remain stationary over one sensing
block, and its reflection coefficient follows the classical Swerling-I
model: $\alpha$ remains constant within a block and varies independently
across blocks according to $\alpha\sim\mathcal{CN}(0,1)$. This model
captures blockwise target-reflection fluctuations while maintaining a
tractable single-target detection model.

Both the transmit and sensing receive arrays are uniform linear arrays
(ULAs) with half-wavelength antenna spacing. The steering vector of an
$N_q$-element array, $q\in\{t,r\}$, is $\mathbf{a}_q(\theta) = \begin{bmatrix}   1 & e^{j\pi\sin\theta} &  \cdots & e^{j\pi(N_q-1)\sin\theta} \end{bmatrix}^{T}$.
Accordingly,
$\|\mathbf{a}_q(\theta)\|^2=N_q$. The corresponding rank-one sensing
channel is modeled as $\mathbf{H}_s(\theta)  = \alpha  \mathbf{a}_r(\theta) \mathbf{a}_t^{H}(\theta)$,
and the sensing received signal is
\begin{equation}
    \mathbf{Y}_s
    =
    \sqrt{\rho_s}\alpha
    \mathbf{a}_r(\theta)
    \mathbf{a}_t^{H}(\theta)\mathbf{X}
    +\mathbf{N}_s,
    \label{eq:sensing_model}
\end{equation}
where $\rho_s$ denotes the sensing SNR before receive-array combining,
and the entries of $\mathbf{N}_s$ follow $\mathcal{CN}(0,1)$.

\subsection{Diversity--Multiplexing Tradeoff}

For the communication link, we characterize the asymptotic
reliability and spectral-efficiency performance through the
DMT \cite{zheng2003diversity}. In a MIMO fading channel,
spatial degrees of freedom can be exploited either to improve
transmission reliability through diversity or to increase
the transmission rate through spatial multiplexing.
The DMT provides the fundamental tradeoff between
these two gains in the high-SNR regime.

Consider a family of space-time codebooks $\mathcal{C}(\rho_c)$ with transmission rate $R(\rho_c)$ at communication SNR $\rho_c$. The multiplexing and diversity gains are defined, respectively, as
\begin{equation}
    r
    \triangleq
    \lim_{\rho_c\rightarrow\infty}
    \frac{R(\rho_c)}{\log \rho_c},
    \qquad
    d(r)
    \triangleq
    -\lim_{\rho_c\rightarrow\infty}
    \frac{\log P_e(\rho_c)}{\log \rho_c},
    \label{eq:dmt_definitions}
\end{equation}
where $P_e(\rho_c)$ denotes the communication error probability under ML decoding.
Equivalently, at high SNR, $P_e(\rho_c)\doteq \rho_c^{-d(r)},$ where $\doteq$ denotes exponential equality with respect to $\rho_c$.
Thus, a larger $d(r)$ implies a faster asymptotic decay of the communication error probability.

For an $N_t\times N_c$ quasi-static i.i.d. Rayleigh fading
channel, the optimal DMT is characterized by a
piecewise-linear tradeoff between diversity and
multiplexing gains. Space-time codes achieving this
tradeoff are referred to as DMT-optimal codes.



\subsection{Single-Target Detection Performance}

Based on the sensing model in
\eqref{eq:sensing_model}, the target-detection problem is formulated as
the binary hypothesis test
\begin{equation}
\begin{cases}
    \mathcal{H}_0:
    \mathbf{Y}_s=\mathbf{N}_s,\\[1mm]
    \mathcal{H}_1:
    \mathbf{Y}_s=
    \sqrt{\rho_s}\alpha
    \mathbf{a}_r(\theta)\mathbf{a}_t^H(\theta)\mathbf{X}
    +\mathbf{N}_s,
\end{cases}
\label{eq:hypothesis_test}
\end{equation}
where $\mathcal{H}_0$ and $\mathcal{H}_1$ denote the absence and
presence of the target, respectively. For subsequent analysis, we define the directional sensing energy of codeword $\mathbf{X}$ toward angle $\theta$ as
\begin{equation}
    \lambda_{\mathbf{X}}(\theta)
    \triangleq
    \mathbf{a}_t^H(\theta)
    \mathbf{X}\mathbf{X}^H
    \mathbf{a}_t(\theta),
    \label{eq:sensing_energy}
\end{equation}
which quantifies the transmitted energy projected onto the steering direction $\theta$. Following classical radar detection theory
\cite{richards2010principles}, marginalizing over the
target coefficient $\alpha$ yields an exponentially
distributed matched-filter statistic with means
$1$ and $1+N_r\rho_s\lambda_{\mathbf{X}}(\theta)$
under $\mathcal{H}_0$ and $\mathcal{H}_1$,
respectively. Setting the detection threshold
according to a prescribed false-alarm probability
$P_{\mathrm{FA}}$, the conditional miss-detection
probability is given by
\begin{equation}
    P_{\mathrm{MD}}(\rho_s|\mathbf{X},\theta)
    =
    1-
    P_{\mathrm{FA}}^{
    \frac{1}{
    1+N_r\rho_s\lambda_{\mathbf{X}}(\theta)}}.
    \label{eq:pmd_closed_form}
\end{equation}

Since both the transmitted codeword and the target direction may vary,
we evaluate the overall sensing performance by averaging over
$\mathbf{X}$ and $\Theta$:
\begin{equation}
    \overline{P}_{\mathrm{MD}}(\rho_s)
    \triangleq
    \mathbb{E}_{\mathbf{X},\Theta}
    \left[
    1-
    P_{\mathrm{FA}}^{
    \frac{1}{
    1+N_r\rho_s\lambda_{\mathbf{X}}(\Theta)}}
    \right].
    \label{eq:expected_pmd}
\end{equation}
We assume equiprobable codewords and
$\Theta\sim\mathcal{U}[-\omega,\omega]$, where
$[-\omega,\omega]$ denotes the angular sensing region of interest.

\section{DMTS of the Perfect Code}
In this section, we consider square space-time codes with
$N_t=T=N$, where each codeword
$\mathbf{X}\in\mathbb{C}^{N\times N}$ spans $N$ channel uses. Using the DMT-optimal Perfect code \cite{oggier2006perfect}, we examine how its algebraic properties affect sensing and reveal limitations over restricted sensing regions.

\subsection{DMT-Optimal Space-Time Codes with NVD}

We focus on full-rate space-time codes satisfying the
non-vanishing-determinant (NVD) property, defined as
\begin{equation}
    \inf_{\Delta\mathbf{X}\neq\mathbf{0}}
    \det\left(
        \Delta\mathbf{X}\Delta\mathbf{X}^{H}
    \right)>0,
    \label{eq:nvd}
\end{equation}
where $\Delta\mathbf{X}$ denotes a nonzero lattice-codeword
difference. The NVD property guarantees full-rank codeword
differences and, together with the full-rate property,
provides a key algebraic criterion for DMT-optimal
constructions \cite{elia2006explicit,oggier2007cyclic}.
The Perfect code
is a representative example
\cite{oggier2006perfect}.

\subsection{Sensing Diversity and Joint DMTS Optimality}
We next characterize the sensing reliability of a
space-time codeword under the single-target detection
model. Motivated by the diversity
viewpoint in radar detection \cite{fishler2006spatial},
we introduce the following sensing diversity order
to characterize the high-SNR decay of the
miss-detection probability.

\begin{definition}[Sensing Diversity]
The sensing diversity order of a transmitted
codeword $\mathbf{X}$ at direction $\theta$
is defined as
\begin{equation}
    d_s(\mathbf{X},\theta)
    \triangleq
    -\lim_{\rho_s\rightarrow\infty}
    \frac{
    \log P_{\mathrm{MD}}(\rho_s|\mathbf{X},\theta)}
    {\log\rho_s}.
    \label{eq:sensing_diversity}
\end{equation}
\end{definition}

From \eqref{eq:pmd_closed_form}, the sensing diversity
depends on the directional sensing gain
$\lambda_{\mathbf{X}}(\theta)$.
In particular, directions with zero sensing gain
are referred to as sensing blind spots.

\begin{definition}[Sensing Blind Spot]
For a transmitted codeword $\mathbf{X}$,
a target direction $\theta$ is referred to as
a sensing blind spot if
$\lambda_{\mathbf{X}}(\theta)=0$.
\end{definition}

At a sensing blind spot, no transmit energy is
projected onto the corresponding target direction.
The following proposition characterizes the
resulting sensing diversity.

\begin{proposition}
\label{prop:sensing_diversity}
For the single-target detection model considered
in Section~II-C, the sensing diversity order
satisfies
\begin{equation}
    d_s(\mathbf{X},\theta)
    =
    \begin{cases}
        1, & \lambda_{\mathbf{X}}(\theta)>0,\\
        0, & \lambda_{\mathbf{X}}(\theta)=0.
    \end{cases}
    \label{eq:sensing_diversity_result}
\end{equation}
Hence, the maximum sensing diversity order
under the considered model is one.
\end{proposition}

\begin{proof}
For $\lambda_{\mathbf{X}}(\theta)>0$,
the high-SNR expansion of
\eqref{eq:pmd_closed_form} yields
\begin{equation}\label{eqn:P_MD_high_SNR}
    P_{\mathrm{MD}}(\rho_s|\mathbf{X},\theta)
    =
    \frac{-\log P_{\mathrm{FA}}}
    {N_r\rho_s\lambda_{\mathbf{X}}(\theta)}
    +O(\rho_s^{-2}),
\end{equation}
which gives $d_s(\mathbf{X},\theta)=1$.
For $\lambda_{\mathbf{X}}(\theta)=0$,
$P_{\mathrm{MD}}=1-P_{\mathrm{FA}}$
is independent of $\rho_s$, yielding
$d_s(\mathbf{X},\theta)=0$.
\end{proof}

The above result reveals two distinct high-SNR sensing behaviors. Positive directional sensing energy yields the maximum sensing diversity order of one, whereas a sensing blind spot results in an irreducible miss-detection floor.

\begin{definition}[Joint DMT--Sensing-Diversity (DMTS) Optimality]
A space-time coding scheme achieves joint DMTS optimality if it attains the optimal communication DMT and maximum sensing diversity for every nonzero codeword, i.e.,
\begin{equation}
    d(r)=d^{\star}(r), \qquad
    d_s(\mathbf{X},\theta)=1,\quad
    \forall\,\mathbf{X}\neq\mathbf{0},\ \forall\,\theta.
\end{equation}
\end{definition}

\begin{theorem}[Joint DMTS Optimality of NVD Space-Time Codes]
\label{thm:nvd_dmts}
Consider a DMT-optimal space-time code family constructed
from a linear lattice containing the zero matrix and
satisfying the NVD property. Then, the code family
achieves joint DMTS optimality for all nonzero codewords.
\end{theorem}

\begin{proof}
Since the underlying lattice contains the zero matrix, applying \eqref{eq:nvd} to $\Delta\mathbf{X}=\mathbf{X}-\mathbf{0}$ ensures that every nonzero codeword is full rank. Hence, $\mathbf{X}\mathbf{X}^H\succ\mathbf{0}$, implying  $\lambda_{\mathbf{X}}(\theta)>0$ for all $\theta$.
Thus, no sensing blind spots exist, and the maximum sensing diversity order of one is achieved. Together with DMT optimality, this establishes joint DMTS optimality.
\end{proof}

As a DMT-optimal NVD space-time code, the Perfect
code achieves joint DMTS optimality according to
Theorem~\ref{thm:nvd_dmts}. However, this asymptotic
optimality does not fully characterize its finite-SNR
sensing performance, which also depends on the
directional sensing gain. We therefore examine the
spatial energy distribution of the Perfect code
in the following subsection.

\subsection{Spatial Sensing Limitation of the Perfect Code}

To examine the spatial energy distribution of the Perfect code, consider its average transmit covariance under independent, zero-mean information symbols with equal average power:
\begin{equation}
    \mathbf{R}_{\mathbf{X}}
    \triangleq
    \mathbb{E}_{\mathbf{X}}
    [\mathbf{X}\mathbf{X}^{H}]
    =
    \mathbf{I}_{N}.
    \label{eq:perfect_covariance}
\end{equation}
Consequently, its average directional sensing energy is $\bar{\lambda}_{\mathrm{PC}}(\theta)
    \triangleq
    \mathbb{E}_{\mathbf{X}}
    [\lambda_{\mathbf{X}}(\theta)]
    =
    \mathbf{a}_{t}^{H}(\theta)
    \mathbf{R}_{\mathbf{X}}
    \mathbf{a}_{t}(\theta)
    =
    N$,
where the last equality follows from the $N$-element ULA model.

While such isotropic energy distribution provides uniform angular coverage, it cannot exploit prior knowledge of a restricted target angular region to concentrate sensing energy. Thus, joint DMTS optimality does not necessarily imply optimal sensing performance, motivating the full-rank spatial precoding framework developed in the next section to reshape directional sensing energy while preserving joint DMTS optimality.

\section{Communication--Sensing Gain Tradeoff under Joint
DMTS Optimality}
To improve the Perfect code’s isotropic sensing performance, we show that fixed full-rank precoding preserves joint DMTS optimality, characterize the tradeoff between communication coding gain and sensing performance, and design precoders exploiting angular priors.

\subsection{Proposed Precoded Space-Time Coding Model}
Consider a DMTS-optimal space-time code $\mathcal{C}$ satisfying the NVD property. We consider full-rank precoders $\mathbf{F}\in\mathbb{C}^{N\times N}$ with $\mathbf{Q}\triangleq\mathbf{F}\mathbf{F}^{H}\succ 0$ and $\operatorname{tr}(\mathbf{Q})=N$. The precoded space-time code is defined as $\mathcal{C}_{\mathbf{F}}\triangleq\{\mathbf{F}\mathbf{X}\mid\mathbf{X}\in\mathcal{C}\}$. We first show that $\mathcal{C}_{\mathbf{F}}$ preserves DMTS optimality.



\begin{theorem}
\label{thm:full_rank_preservation}
For any DMTS-optimal space-time code $\mathcal{C}$ satisfying NVD and any full-rank precoder $\mathbf{F}\in\mathbb{C}^{N\times N}$, the precoded space-time code $\mathcal{C}_{\mathbf{F}}\triangleq \{\mathbf{FX}|\mathbf{X}\in \mathcal{C}\}$ preserves joint DMTS optimality. i.e.,
$    d_{\mathbf{F}}(r)=d^{\star}(r),~
    d_s(\mathbf{F}\mathbf{X},\theta)=1,
$
for all admissible $r$, nonzero codewords $\mathbf{X}$, and sensing directions $\theta$.
\end{theorem}

\begin{proof}[Proof sketch]
Since $\mathbf{F}$ is fixed and full rank, precoding scales every
nonzero codeword-difference determinant by the positive constant
$\det(\mathbf{F}\mathbf{F}^{H})$, and hence preserves the NVD
property. Moreover, the singular values of $\mathbf{F}$ are bounded
away from zero and infinity by SNR-independent constants. The
instantaneous capacity of the precoded channel is therefore bounded
between those of the unprecoded channel with two fixed SNR scalings,
which preserves the multiplexing gain and hence the optimal DMT.
Finally, since both $\mathbf{F}$ and every nonzero NVD codeword are
full rank, $\mathbf{F}\mathbf{X}$ remains full rank and produces no
sensing blind spots. By Proposition~\ref{prop:sensing_diversity}, the
maximum sensing diversity order of one is therefore preserved.
\end{proof}

Theorem~2 formalizes the role of full-rank precoding in the proposed
space-time-coded ISAC framework. Unlike channel-adaptive low-rank
signaling, full-rank precoding retains the channel-independent DMT
guarantee of the underlying space-time code while introducing
additional spatial degrees of freedom for reshaping the sensing gain.

\subsection{Communication--Sensing Gain Tradeoff}
Theorem~\ref{thm:full_rank_preservation} establishes the joint DMTS optimality for full rank precoders. However, they are relevant when $\rho_c$ is asymptotically large. Here, we discuss the communication and sensing coding gains in the finite $\rho_c$ regime.
For full-diversity algebraic space-time codes, the minimum determinant is a standard measure of the high-SNR coding gain \cite{oggier2007cyclic}. Specifically, define the minimum determinant
of the original code as $\delta_{\min}
    \triangleq
    \inf_{\Delta\mathbf{X}\neq\mathbf{0}}
    \det\left(
        \Delta\mathbf{X}\Delta\mathbf{X}^{H}
    \right)$.
After precoding, the minimum determinant becomes
\begin{align}
    \widetilde{\delta}_{\min}
    &=
    \inf_{\Delta\mathbf{X}\neq\mathbf{0}}
    \det\left[
        (\mathbf{F}\Delta\mathbf{X})
        (\mathbf{F}\Delta\mathbf{X})^{H}
    \right]
    \nonumber =
    \det(\mathbf{Q})\delta_{\min},
    \label{eq:precoded_min_det}
\end{align}
The arithmetic-geometric mean inequality gives $\det(\mathbf{Q})\leq 1$, with equality when $\mathbf{Q}=\mathbf{I}_{N}$. 
We define the communication coding-gain backoff relative to the unprecoded space-time code as
\begin{equation}
    B_c(\mathbf{F})
    \triangleq
    -\frac{10}{N}
    \log_{10}
    \det\left(
        \mathbf{F}\mathbf{F}^{H}
    \right)
    \geq 0 .
    \label{eq:comm_backoff}
\end{equation}
Here, $B_c(\mathbf{I}_N)=0$, and a larger $B_c(\mathbf{F})$ indicates a greater reduction in communication coding gain permitted to improve sensing performance.

On the sensing side, define
\begin{equation}
    J(\mathbf{F})
    \triangleq
    \mathbb{E}_{\mathbf{X},\Theta}
    \left[
        \frac{1}{
        \lambda_{\mathbf{F},\mathbf{X}}(\Theta)}
    \right].
    \label{eq:sensing_objective}
\end{equation}
For a full-rank precoder, averaging \eqref{eqn:P_MD_high_SNR} leads to 
\begin{equation}
    \overline{P}_{\mathrm{MD}}(\rho_s)
    =
    \frac{-\log P_{\mathrm{FA}}}{N\rho_s}
    J(\mathbf{F})
    +
    O(\rho_s^{-2}).
    \label{eq:high_snr_pmd}
\end{equation}
Therefore, a smaller $J(\mathbf{F})$ corresponds to a larger
high-SNR sensing gain. To quantify the sensing improvement relative to the unprecoded 
space-time code, we define
\begin{equation}
    G_s(\mathbf{F})
    \triangleq
    10\log_{10}
    \frac{
        J(\mathbf{I}_{N})
    }{
        J(\mathbf{F})
    } .
    \label{eq:sensing_gain}
\end{equation}
Hence, $G_s(\mathbf{I}_{N})=0$, while
$G_s(\mathbf{F})>0$ indicates a high-SNR sensing-gain improvement over
the original 
space-time code.

For a prescribed communication coding-gain backoff budget
$B_{\max}$, the precoder design problem is formulated as
\begin{flalign*}
&\textbf{(P1)}\quad
\begin{aligned}[t]
    \min_{\mathbf{F}}
    \quad &
    J(\mathbf{F})\\
    \operatorname{s.t.}
    \quad &
    \operatorname{tr}
    \left(\mathbf{F}\mathbf{F}^{H}\right)=N, ~ B_c(\mathbf{F})\leq B_{\max},~
    \operatorname{rank}(\mathbf{F})=N.
\end{aligned}
&&
\end{flalign*}
By varying $B_{\max}$, problem \textbf{(P1)} characterizes the
tradeoff between communication coding-gain backoff and
sensing performance improvement relative to the unprecoded
Perfect code, while preserving joint DMTS optimality.

\subsection{Precoder Optimization}
In what follows, we focus on a class of NVD-achieving codes, the Perfect code $\mathcal{C}_\mathrm{PC}=\{\mathbf{X}_{\mathrm{PC}}\}$, as a representative.
We first note that \textbf{(P1)} is generally challenging to solve, because the full-rank requirement on $\mathbf{F}$ leads to a non-convex feasible set. We thus apply Jensen's inequality to the objective function     
\begin{equation}
J(\mathbf{F})
    \overset{(a)}{\geq}
    \mathbb{E}_{\Theta}
    \left[
        \frac{1}{
        \mathbb{E}_{\mathbf{X}}
        [\lambda_{\mathbf{X}}(\Theta)]}
    \right]
    \overset{(b)}{=}
    \mathbb{E}_{\Theta}
    \left[
        \frac{1}{
        \mathbf{a}_{t}^{H}(\Theta)
        \mathbf{Q}
        \mathbf{a}_{t}(\Theta)}
    \right], \label{eqn:Objective_Jensen}
\end{equation}
where (a) is due to the convexity of $f(x)=1/x$ and (b) uses the fact that $\mathbb{E}_{\mathbf{X}}[\mathbf{X}\mathbf{X}^{H}] =\mathbf{I}_{N}$ for the Perfect code.

This motivates the Jensen-based covariance design
\begin{equation*}
\textbf{(P2)}\quad
\begin{aligned}[t]
    \underset{\mathbf{Q}\succ0}{\operatorname{min}}
    \quad &
    \mathbb{E}_{\Theta}
    \left[
        \frac{1}{
        \mathbf{a}_{t}^{H}(\Theta)
        \mathbf{Q}
        \mathbf{a}_{t}(\Theta)}
    \right]\\
    \operatorname{s.t.}
    \quad &
    \operatorname{tr}(\mathbf{Q})=N,\quad
    \log\det(\mathbf{Q})
    \geq
    -\frac{N\ln 10}{10}B_{\max}.
\end{aligned}
\end{equation*}
\textbf{(P2)} is a convex optimization problem because its objective function is convex in $\mathbf{Q}$, and the full-rank constraint on $\mathbf{F}$ is equivalently expressed as $\mathbf{Q}=\mathbf{F}\mathbf{F}^{H}\succ0$, which defines a convex feasible set. Once the optimized $\mathbf{Q}^{\star}$ is obtained, eigenvale decomposition can be employed to obtain its principle square root as precoder $\mathbf{F}^{\star}$.


To obtain a more interpretable low-complexity design, we further consider a surrogate that maximizes the average sensing gain over the angular region. That is, we maximize $\mathbb{E}_{\Theta}  \left[ \mathbf{a}_{t}^{H}(\Theta)   \mathbf{Q}   \mathbf{a}_{t}(\Theta)  \right]  = \operatorname{tr} (\mathbf{R}_{\Theta}\mathbf{Q})$, where $\mathbf{R}_{\Theta} \triangleq  \mathbb{E}_{\Theta} \left[    \mathbf{a}_{t}(\Theta)   \mathbf{a}_{t}^{H}(\Theta) \right]$, which leads us to the surrogate problem:

\begin{equation*}
\textbf{(P3)}\quad
\begin{aligned}[t]
    \underset{\mathbf{Q}\succ0}{\operatorname{max}}
    \quad &
    \operatorname{tr}(\mathbf{R}_{\Theta}\mathbf{Q})\\
    \operatorname{s.t.}
    \quad &
    \operatorname{tr}(\mathbf{Q})=N, \quad\log\det(\mathbf{Q})
    \geq
    -\frac{N\ln 10}{10}B_{\max}.
\end{aligned}
\end{equation*}

\begin{proposition}
\label{prop:surrogate_solution}
The optimal solution to {\bf (P3)} is $\mathbf{Q}_{\mathrm{S}}^{\star}=\mathbf{I}$ if $\mathbf{R}_{\Theta}=\alpha\mathbf{I}$ for some constant $\alpha$ and
\begin{equation}
    \mathbf{Q}_{\mathrm{S}}^{\star}(\mu)
    =
    \frac{
        N(\mu\mathbf{I}_{N}-\mathbf{R}_{\Theta})^{-1}
    }{
        \operatorname{tr}
        \left[
            (\mu\mathbf{I}_{N}-\mathbf{R}_{\Theta})^{-1}
        \right]
    },\quad \text{otherwise,}
    \label{eq:surrogate_solution}
\end{equation}
where $\mu>\lambda_{\max}(\mathbf{R}_{\Theta})$ is chosen such that the determinant constraint of problem \textbf{(P3)} holds with equality.
\end{proposition}

\begin{proof}[Proof sketch]
For $\mathbf{R}_{\Theta}=\alpha\mathbf{I}$, it is trivial. Consider $\mathbf{R}_{\Theta}\neq \alpha\mathbf{I}$. Applying the KKT conditions yields $-\mathbf{R}_{\Theta} +\mu\mathbf{I}_N -\nu\mathbf{Q}^{-1}  =\mathbf{0}$ and $\nu>0$, which
implies 
$\mathbf{Q}\propto
(\mu\mathbf{I}_N-\mathbf{R}_{\Theta})^{-1}$
with $\mu>\lambda_{\max}(\mathbf{R}_{\Theta})$.
The trace constraint is then used to determine the scaling factor, while complementary slackness ensures that the determinant constraint is active.
\end{proof}

The three optimization problems provide successive simplifications of the original precoder design. From \textbf{(P1)} to \textbf{(P2)}, Jensen's inequality is applied over the transmitted codewords, with equality when $\mathbf{X}_{\mathrm{PC}}\mathbf{X}_{\mathrm{PC}}^H$ is deterministic. From \textbf{(P2)} to \textbf{(P3)}, a second application of Jensen's inequality over the target angle yields a lower bound that is tight when the average directional sensing energy is constant throughout the angular region. Consequently, the surrogate design is expected to closely approximate the Jensen-based optimization when the directional sensing energy varies little over a narrow sensing region.

The solution in \eqref{eq:surrogate_solution} further illustrates the communication--sensing gain tradeoff. When $B_{\max}=0$, the determinant and power constraints enforce $\mathbf{Q}=\mathbf{I}_N$, recovering the unprecoded Perfect code. As $B_{\max}$ increases, more spatial energy can be allocated toward the dominant eigendirections of $\mathbf{R}_{\Theta}$ while preserving joint DMTS optimality. Thus, communication coding-gain backoff enables improved sensing performance without sacrificing asymptotic optimality. While {\bf (P2)} provides an optimistic benchmark for the original problem, {\bf (P3)} admits a simple closed-form structure based on the angular prior.

\section{Simulation Results}
We evaluate the proposed full-rank precoding framework using the $4\times4$ Perfect code over the quasi-static i.i.d. Rayleigh channel. The target direction follows $\Theta\sim\mathcal{U}[-\omega,\omega]$, with $\omega\in\{\pi/10,\pi/5\}$ illustrating the impact of directional priors. All gains are relative to the unprecoded Perfect code.

\subsection{Communication--Sensing Gain Tradeoff}

\begin{figure}[!h]
    \centering
    \vspace{-15pt}
    \subfloat[$\omega=\pi/10$]{
        \includegraphics[width=0.47\columnwidth]{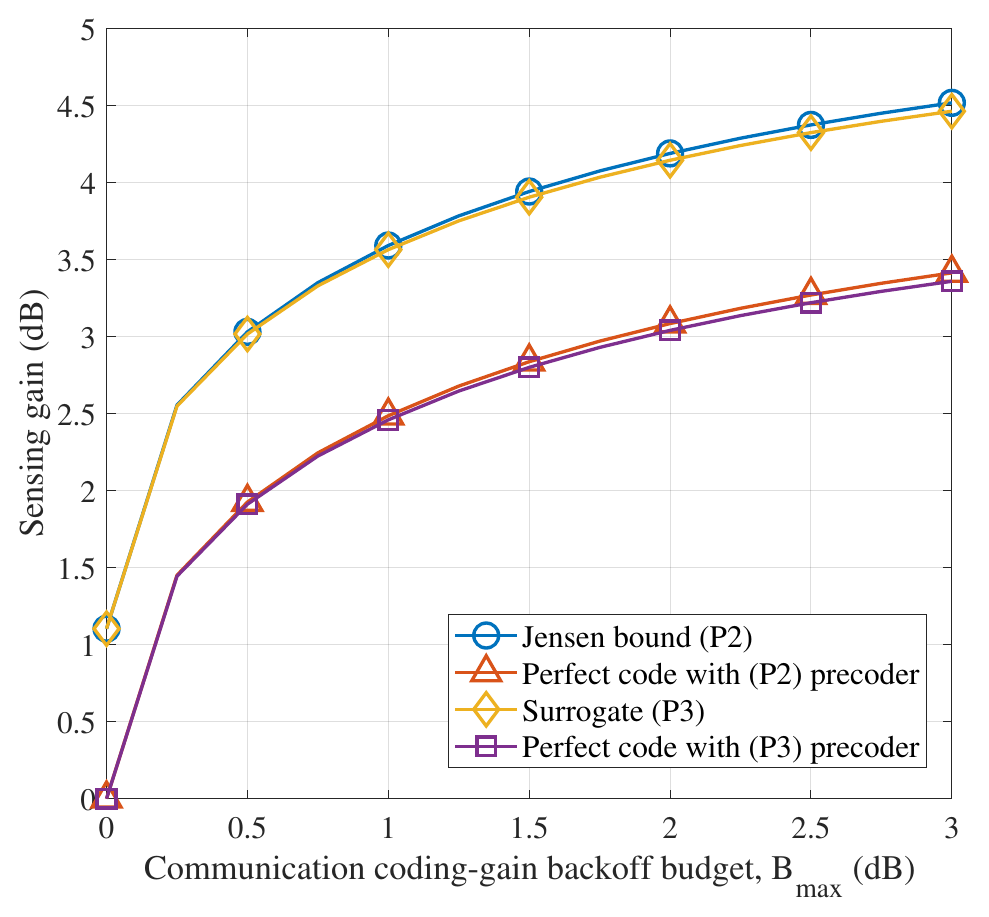}
    }
    \hfill
    \subfloat[$\omega=\pi/5$]{
        \includegraphics[width=0.47\columnwidth]{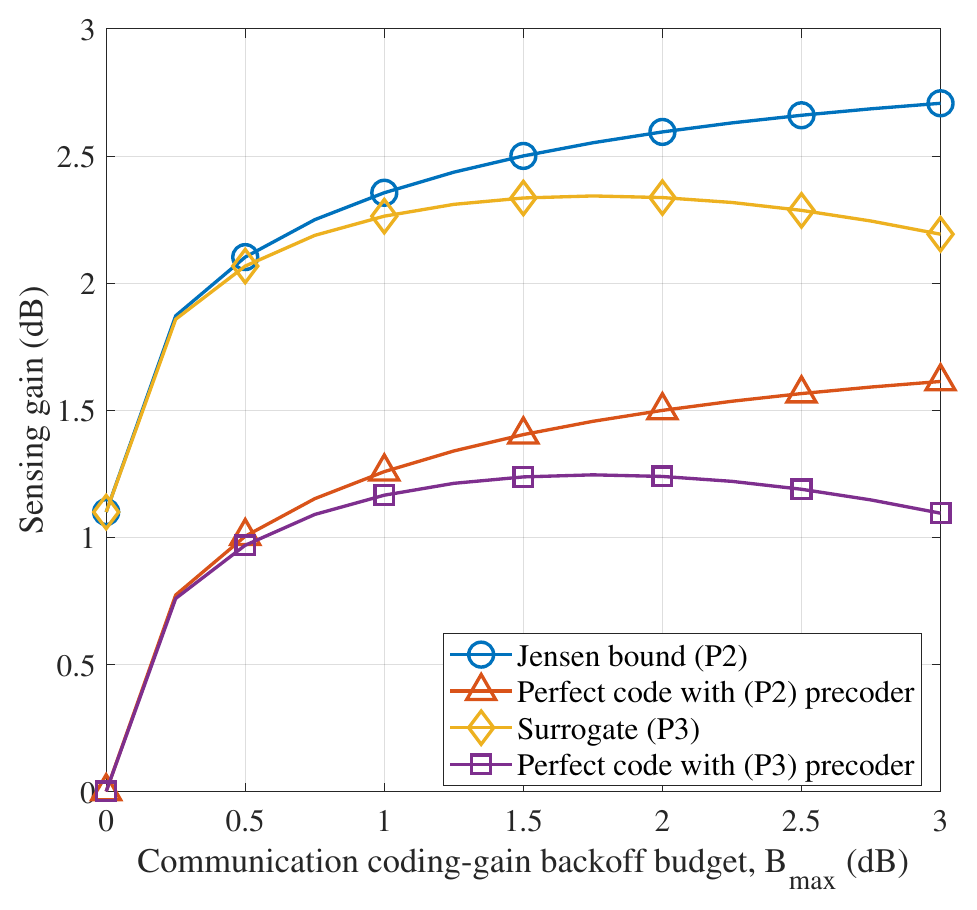}
    }
    \caption{Communication--sensing gain tradeoff for
    $\omega=\pi/10$ and $\omega=\pi/5$.}
    \label{fig:tradeoff}
\end{figure}

Fig.~\ref{fig:tradeoff} shows the sensing-performance improvement versus the communication coding-gain backoff budget $B_{\max}$. For the narrow sensing region $\omega=\pi/10$, the surrogate design closely follows the Jensen-based design over the entire range of $B_{\max}$, and the corresponding precoded Perfect codes achieve nearly identical sensing performance. At $B_{\max}=3$ dB, the proposed precoder provides approximately $3.4$ dB of sensing-gain improvement over the unprecoded Perfect code.

As the sensing region widens to $\omega=\pi/5$, however, the surrogate gradually deviates from the Jensen-based design as $B_{\max}$ increases. This difference arises because the surrogate maximizes the average directional sensing energy, whereas the original high-SNR objective depends on the average reciprocal directional sensing energy and therefore penalizes low-energy directions more strongly.

\begin{figure}[!h]
    \centering
    \subfloat[$\omega=\pi/10$]{
        \includegraphics[width=0.47\columnwidth]{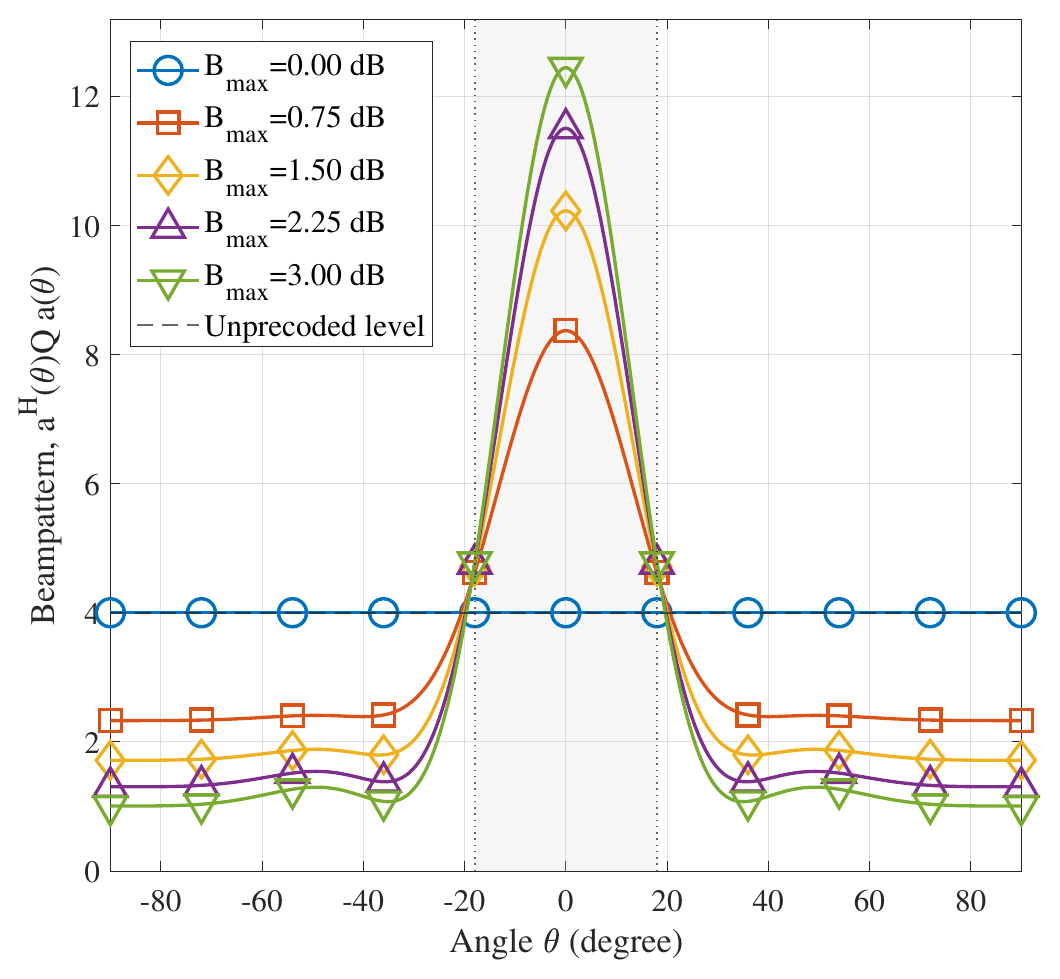}
    }
    \hfill
    \subfloat[$\omega=\pi/5$]{
        \includegraphics[width=0.47\columnwidth]{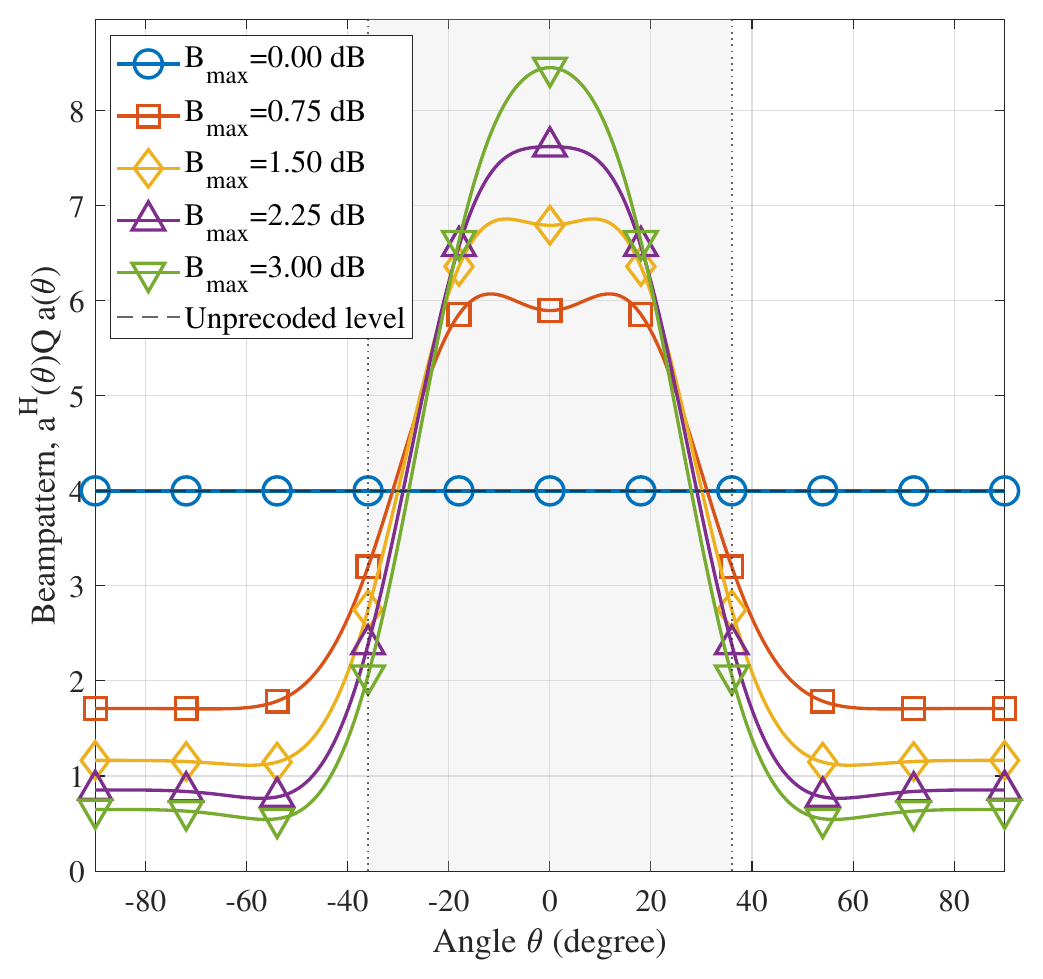}
    }
    \caption{The surrogate-precoder beampattern for
    $\omega=\pi/10$ and $\omega=\pi/5$.}
    \label{fig:beampattern}
\end{figure}

The beampatterns in Fig.~\ref{fig:beampattern} provide a direct interpretation of this behavior. For $\omega=\pi/10$, increasing $B_{\max}$ concentrates spatial energy within the narrow region of interest while maintaining sufficiently high directional sensing energy throughout the region, making the surrogate a close approximation to the Jensen-based objective. For $\omega=\pi/5$, the surrogate increasingly concentrates energy near the center of the wider angular region while reducing the directional sensing energy near its boundaries. Although this can increase the average directional sensing energy, low-energy directions are strongly penalized by the reciprocal-energy objective, causing the surrogate to lose accuracy at larger $B_{\max}$. This also explains why the surrogate design is particularly effective when the available angular prior is narrow.

\begin{figure}[!h]
    \centering
    \subfloat[$\omega=\pi/10$]{
        \includegraphics[width=0.47\columnwidth]{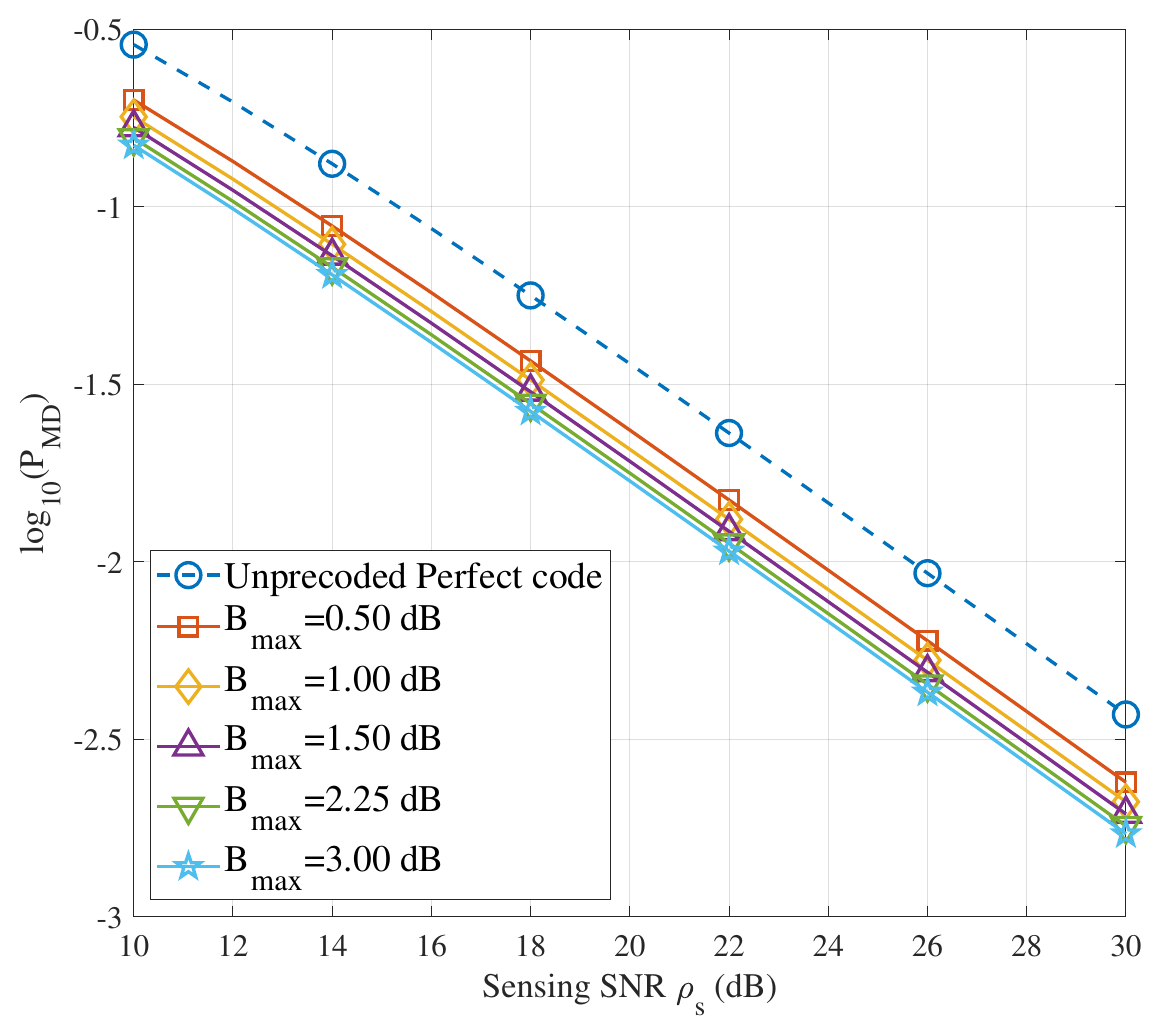}
    }
    \hfill
    \subfloat[$\omega=\pi/5$]{
        \includegraphics[width=0.47\columnwidth]{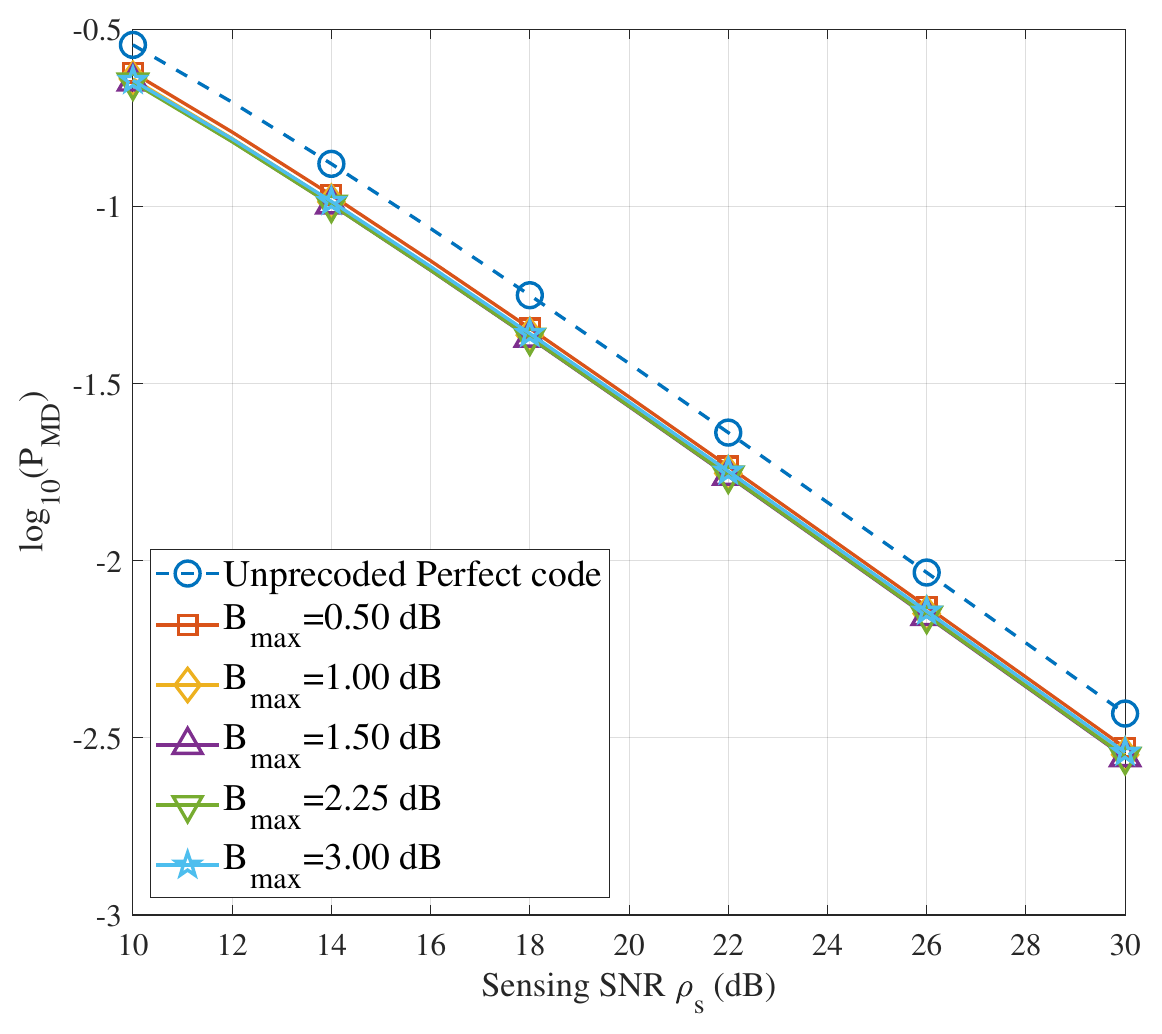}
    }
    \caption{Miss-detection performance of the unprecoded and
    precoded Perfect codes for different communication coding-gain
    loss budgets $B_{\max}$.}
    \vspace{-15pt}
    \label{fig:sensing_performance}
\end{figure}

\begin{figure}[!h]
    \centering
    \subfloat[$\omega=\pi/10$]{
        \includegraphics[width=0.47\columnwidth]{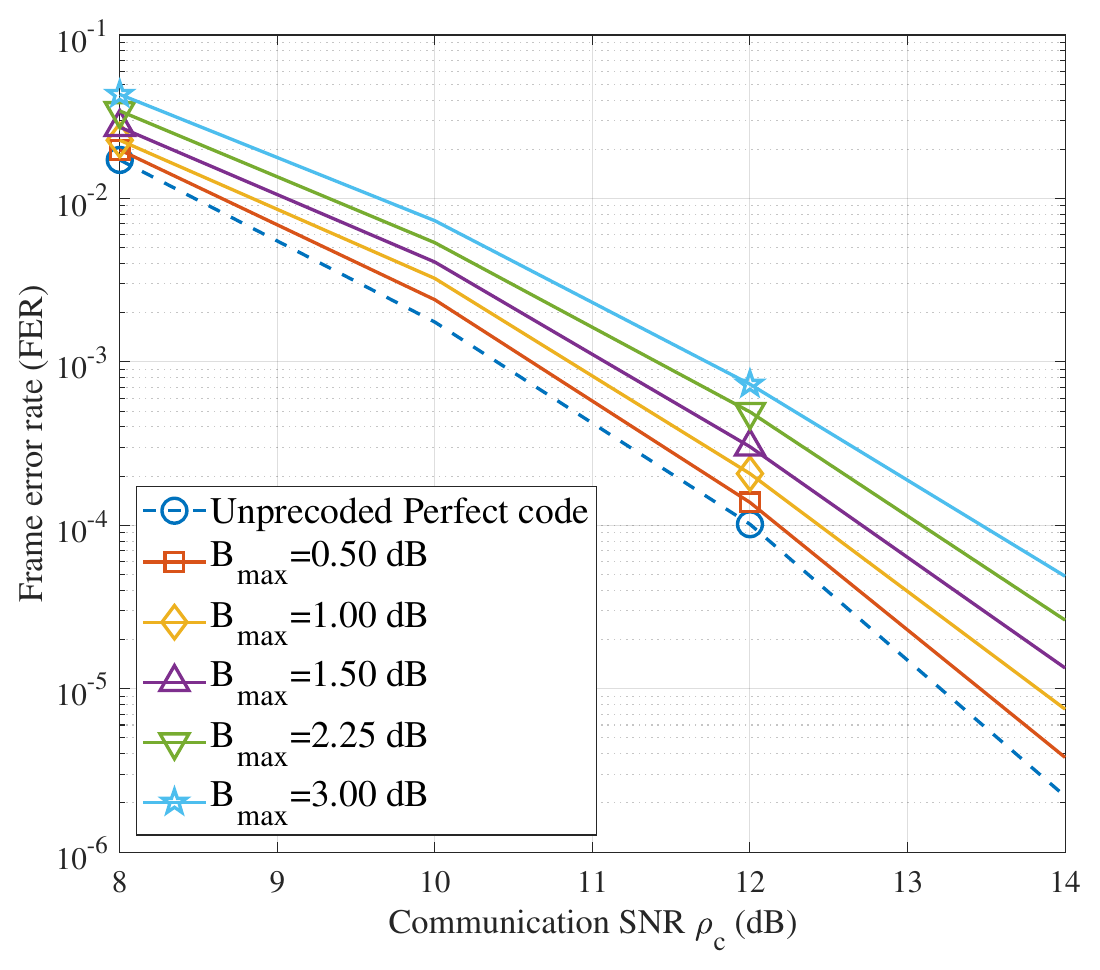}
    }
    \hfill
    \subfloat[$\omega=\pi/5$]{
        \includegraphics[width=0.47\columnwidth]{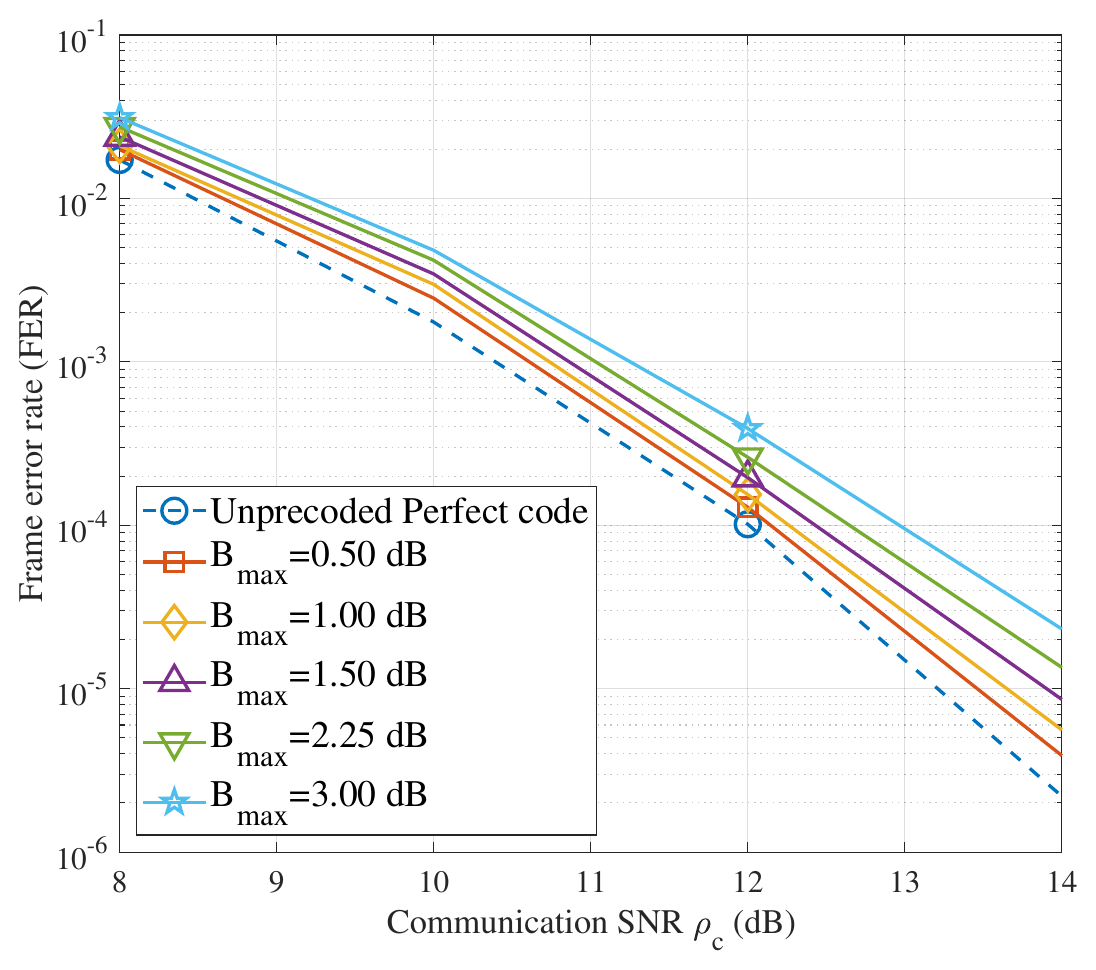}
    }
    \caption{Communication frame error rate of the unprecoded and
    precoded Perfect codes for different communication coding-gain
    loss budgets $B_{\max}$.}
\vspace{-15pt}\label{fig:communication_performance}
\end{figure}

\subsection{Sensing and Communication Performance}

Figs.~\ref{fig:sensing_performance} and \ref{fig:communication_performance} illustrate the sensing and communication performance under different coding-gain backoff budgets $B_{\max}$ using the surrogate precoder {\bf (P3)}.  
 As $B_{\max}$ increases, the miss-detection probability decreases while maintaining a similar high-SNR slope, consistent with the preservation of sensing diversity. The improvement is more pronounced for $\omega=\pi/10$ than for $\omega=\pi/5$, in agreement with Fig.~\ref{fig:tradeoff}.

Meanwhile, Fig.~\ref{fig:communication_performance} shows that increasing $B_{\max}$ degrades the frame error rate (FER) due to reduced determinant-based coding gain. Nevertheless, the precoded schemes exhibit similar high-SNR decay behavior, consistent with the DMT-preserving property established in Section~IV-B. These results demonstrate that controlled communication coding-gain backoff improves sensing performance while preserving joint DMTS optimality.

\section{Conclusion}

In this work, we developed an analytical framework for characterizing the sensing behavior of DMT-optimal space-time codes in ISAC systems. Under a single-target detection model, we showed that the absence of sensing blind spots guarantees the maximum sensing diversity order of one, and established that DMT-optimal NVD space-time codes achieve joint DMTS optimality. Using the Perfect code as a representative example, we identified the sensing limitations arising from its isotropic spatial energy distribution. To address this limitation, we developed a full-rank spatial precoding framework that preserves joint DMTS optimality while enabling a controlled tradeoff between communication coding-gain backoff and sensing performance. Furthermore, the proposed low-complexity surrogate design closely approaches the Jensen-based benchmark for narrow sensing regions.

Future work includes extending the analysis to more general sensing models, where higher sensing-diversity orders and nontrivial DMTS tradeoffs may emerge. Investigating statistically correlated communication and sensing channels, as well as directly exploiting the algebraic structure of space-time codes for joint communication and sensing optimization, also represents promising research directions.

\bibliographystyle{IEEEtran}
\bibliography{references_abbr}
\clearpage

\appendices

\section{Derivation of the Miss-Detection Probability}
\label{app:pmd}

Define the sensing signature as
\begin{equation}
    \mathbf{s}(\mathbf{X},\theta)
    \triangleq
    \operatorname{vec}
    \left(
        \mathbf{a}_r(\theta)
        \mathbf{a}_t^H(\theta)\mathbf{X}
    \right).
    \label{eq:app_sensing_signature}
\end{equation}
Since $\|\mathbf{a}_r(\theta)\|^2=N_r$, we have
\begin{equation}
    \|\mathbf{s}(\mathbf{X},\theta)\|^2
    =
    N_r\lambda_{\mathbf{X}}(\theta).
    \label{eq:app_s_signature_norm}
\end{equation}

For $\lambda_{\mathbf{X}}(\theta)>0$, marginalizing over
$\alpha\sim\mathcal{CN}(0,1)$ reduces the
likelihood-ratio test to a threshold test on the
normalized matched-filter statistic
\begin{equation}
    T
    \triangleq
    \frac{
        \left|
        \mathbf{s}^H(\mathbf{X},\theta)
        \operatorname{vec}(\mathbf{Y}_s)
        \right|^2
    }{
        \|\mathbf{s}(\mathbf{X},\theta)\|^2
    }.
    \label{eq:app_test_statistic}
\end{equation}
Under the two hypotheses,
\begin{align}
    T|\mathcal{H}_0
    &\sim \operatorname{Exp}(1),\\
    T|\mathcal{H}_1
    &\sim \operatorname{Exp}
    \left(
        1+N_r\rho_s
        \lambda_{\mathbf{X}}(\theta)
    \right),
\end{align}
where the argument denotes the mean.

For a detection threshold $\gamma$, the false-alarm
and miss-detection probabilities are
\begin{align}
    P_{\mathrm{FA}}
    &=
    \Pr(T>\gamma|\mathcal{H}_0)
    =e^{-\gamma},\\
    P_{\mathrm{MD}}
    &=
    \Pr(T\leq\gamma|\mathcal{H}_1)
    \nonumber\\
    &=
    1-\exp\left(
        -\frac{\gamma}{
        1+N_r\rho_s\lambda_{\mathbf{X}}(\theta)}
    \right).
\end{align}
Substituting $\gamma=-\log P_{\mathrm{FA}}$
yields \eqref{eq:pmd_closed_form}.
When $\lambda_{\mathbf{X}}(\theta)=0$, the two
hypotheses are indistinguishable, giving
$P_{\mathrm{MD}}=1-P_{\mathrm{FA}}$,
which is also consistent with
\eqref{eq:pmd_closed_form}.

\section{Proof of Proposition~\ref{prop:sensing_diversity}}
\label{app:sensing_diversity}

Let $c\triangleq-\log P_{\mathrm{FA}}>0$.
For $\lambda_{\mathbf{X}}(\theta)>0$,
\eqref{eq:pmd_closed_form} gives
\begin{align}
    P_{\mathrm{MD}}(\rho_s|\mathbf{X},\theta)
    &=
    1-\exp\left(
        -\frac{c}{
        1+N_r\rho_s\lambda_{\mathbf{X}}(\theta)}
    \right)
    \nonumber\\
    &=
    \frac{c}{
        N_r\rho_s\lambda_{\mathbf{X}}(\theta)}
    +O(\rho_s^{-2}),
\end{align}
as $\rho_s\rightarrow\infty$. Consequently,
\begin{equation}
    d_s(\mathbf{X},\theta)
    =
    -\lim_{\rho_s\rightarrow\infty}
    \frac{
        \log P_{\mathrm{MD}}(\rho_s|\mathbf{X},\theta)
    }{
        \log\rho_s
    }
    =1.
\end{equation}

If $\lambda_{\mathbf{X}}(\theta)=0$, then
$P_{\mathrm{MD}}=1-P_{\mathrm{FA}}$,
which is independent of $\rho_s$.
Hence, $d_s(\mathbf{X},\theta)=0$,
completing the proof.

\section{Proof of Theorem~\ref{thm:full_rank_preservation}}
\label{app:full_rank}

We first consider communication DMT preservation.
For any nonzero codeword difference
$\Delta\mathbf{X}=\mathbf{X}_1-\mathbf{X}_2$,
the corresponding precoded difference satisfies
\begin{align}
    &\det\left[
        (\mathbf{F}\Delta\mathbf{X})
        (\mathbf{F}\Delta\mathbf{X})^H
    \right]
    \nonumber\\
    &\qquad=
    \det(\mathbf{Q})
    \det\left(
        \Delta\mathbf{X}\Delta\mathbf{X}^H
    \right).
\end{align}
Since $\mathbf{F}$ is fixed and full rank,
$\det(\mathbf{Q})>0$ is independent of SNR.
Therefore, the underlying NVD property is
preserved up to a positive constant factor.
Moreover, full-rank precoding does not change
the code rate.

To examine the channel behavior, note that
\begin{equation}
    \sigma_{\min}^2(\mathbf{F})\mathbf{I}_N
    \preceq
    \mathbf{Q}
    \preceq
    \sigma_{\max}^2(\mathbf{F})\mathbf{I}_N,
\end{equation}
where both singular-value bounds are positive,
finite, and independent of SNR. Consequently,
\begin{align}
    &\log\det\left(
        \mathbf{I}_{N_c}
        +\rho_c\sigma_{\min}^2(\mathbf{F})
        \mathbf{H}_c\mathbf{H}_c^H
    \right)
    \nonumber\\
    &\leq
    \log\det\left(
        \mathbf{I}_{N_c}
        +\rho_c\mathbf{H}_c
        \mathbf{Q}\mathbf{H}_c^H
    \right)
    \nonumber\\
    &\leq
    \log\det\left(
        \mathbf{I}_{N_c}
        +\rho_c\sigma_{\max}^2(\mathbf{F})
        \mathbf{H}_c\mathbf{H}_c^H
    \right).
\end{align}
The bounds differ from the unprecoded
channel only by constant SNR scaling,
which does not affect the outage DMT.
Together with the preserved full-rate
NVD structure, this establishes
\begin{equation}
    d_{\mathbf{F}}(r)=d^\star(r).
\end{equation}

We next consider sensing diversity.
Since the underlying NVD lattice contains
the zero matrix, every nonzero codeword
$\mathbf{X}$ is full rank.
As $\mathbf{F}$ is also full rank,
\begin{equation}
    \mathbf{F}\mathbf{X}\mathbf{X}^H
    \mathbf{F}^H\succ\mathbf{0}.
\end{equation}
Hence, for every sensing direction $\theta$,
\begin{equation}
    \lambda_{\mathbf{F}\mathbf{X}}(\theta)
    =
    \mathbf{a}_t^H(\theta)
    \mathbf{F}\mathbf{X}\mathbf{X}^H
    \mathbf{F}^H\mathbf{a}_t(\theta)
    >0.
\end{equation}
Thus, full-rank precoding introduces
no sensing blind spots. By
Proposition~\ref{prop:sensing_diversity},
\begin{equation}
    d_s(\mathbf{F}\mathbf{X},\theta)=1,
    \qquad \forall\,\theta.
\end{equation}
Therefore, the precoded code preserves
joint DMTS optimality.

\section{Proof of Proposition~\ref{prop:surrogate_solution}}
\label{app:surrogate}

For convenience, define
\begin{equation}
    c_{\max}
    \triangleq
    \frac{N\ln 10}{10}B_{\max}.
\end{equation}
The determinant constraint in problem (P3)
is equivalently written as
\begin{equation}
    -\log\det(\mathbf{Q})-c_{\max}\leq0.
\end{equation}
The corresponding Lagrangian is
\begin{align}
    \mathcal{L}(\mathbf{Q},\mu,\nu)
    &=
    -\operatorname{tr}
    (\mathbf{R}_{\Theta}\mathbf{Q})
    +\mu\left(
        \operatorname{tr}(\mathbf{Q})-N
    \right)
    \nonumber\\
    &\quad
    +\nu\left(
        -\log\det(\mathbf{Q})-c_{\max}
    \right),
\end{align}
where $\mu\in\mathbb{R}$ and $\nu\geq0$
are the multipliers associated with the
trace and determinant constraints.

The KKT stationarity condition gives
\begin{equation}
    -\mathbf{R}_{\Theta}
    +\mu\mathbf{I}_N
    -\nu\mathbf{Q}^{-1}
    =\mathbf{0}.
    \label{eq:app_surrogate_stationarity}
\end{equation}
Since $\mathbf{R}_{\Theta}$ is not
proportional to the identity matrix,
$\nu>0$; otherwise, stationarity would
require $\mathbf{R}_{\Theta}=\mu\mathbf{I}_N$.
Therefore,
\begin{equation}
    \mathbf{Q}
    =
    \nu
    (\mu\mathbf{I}_N-\mathbf{R}_{\Theta})^{-1},
\end{equation}
where
$\mu>\lambda_{\max}(\mathbf{R}_{\Theta})$
follows from $\mathbf{Q}\succ\mathbf{0}$.

Enforcing $\operatorname{tr}(\mathbf{Q})=N$
determines
\begin{equation}
    \nu
    =
    \frac{N}{
        \operatorname{tr}\left[
        (\mu\mathbf{I}_N-\mathbf{R}_{\Theta})^{-1}
        \right]
    },
\end{equation}
which yields the solution in
\eqref{eq:surrogate_solution}.

Finally, since $\nu>0$,
complementary slackness requires
\begin{equation}
    \log\det
    \left(
        \mathbf{Q}_{\mathrm{S}}^\star(\mu)
    \right)
    =
    -\frac{N\ln 10}{10}B_{\max}.
\end{equation}
Thus, $\mu$ can be determined through
a one-dimensional search, completing
the proof.

\end{document}